

\def\Titlehh#1#2{\nopagenumbers\abstractfont\hsize=\hstitle\rightline{#1}%
\vskip .2in\centerline{\titlefont #2}\abstractfont\vskip .2in\pageno=0}
\def\newsecnpb#1{\global\advance\secno by1\message{(\the\secno. #1)}
\global\subsecno=0\eqnres@t\centerline{\bf\the\secno. #1}
\writetoca{{\secsym} {#1}}\par\nobreak\medskip\nobreak}
\def\eqnres@t{\xdef\secsym{\the\secno.}\global\meqno=1\bigbreak\bigskip}
\def\sequentialequations{\def\eqnres@t{\bigbreak}}\xdef\secsym{}
\global\newcount\subsecno \global\subsecno=0
\def\subsecnpb#1{\global\advance\subsecno by1
\message{(\secsym\the\subsecno. #1)}
\ifnum\lastpenalty>9000\else\bigbreak\fi
\noindent{\secsym\the\subsecno. #1}\writetoca{\string\quad
{\secsym\the\subsecno.} {#1}}\par\nobreak\medskip\nobreak}

\def\CTPa{\it Center for Theoretical Physics, Department of Physics,
      Texas A\&M University}
\def\CTPb{\it College Station, TX 77843-4242, USA}
\def\HARCa{\it Astroparticle Physics Group,
Houston Advanced Research Center (HARC)}
\def\HARCb{\it The Woodlands, TX 77381, USA}

\def\CERN{\it CERN Theory Division, 1211 Geneva 23, Switzerland}
\def\ie{\hbox{\it i.e.}}     
\def\eg{\hbox{\it e.g.}}

\catcode`\@=11 

\def\lsim{\mathrel{\mathpalette\@versim<}}
\def\gsim{\mathrel{\mathpalette\@versim>}}
\def\@versim#1#2{\vcenter{\offinterlineskip
    \ialign{$\m@th#1\hfil##\hfil$\crcr#2\crcr\sim\crcr } }}
\def\boxit#1{\vbox{\hrule\hbox{\vrule\kern3pt
      \vbox{\kern3pt#1\kern3pt}\kern3pt\vrule}\hrule}}

\def\t1{{\tilde 1}}

\def\JL{J. L. Lopez}
\def\DVN{D. V. Nanopoulos}

\def\GeV{\,{\rm GeV}}
\def\TeV{\,{\rm TeV}}
\def\y{\,{\rm y}}
\def\pb{\,{\rm pb}}
\def\ipb{\,{\rm pb^{-1}}}

\def\NPB#1#2#3{Nucl. Phys. B {\bf#1} (19#2) #3}
\def\PLB#1#2#3{Phys. Lett. B {\bf#1} (19#2) #3}

\def\PRD#1#2#3{Phys. Rev. D {\bf#1} (19#2) #3}
\def\PRL#1#2#3{Phys. Rev. Lett. {\bf#1} (19#2) #3}

\def\MODA#1#2#3{Mod. Phys. Lett. A {\bf#1} (19#2) #3}

\def\TAMU#1{Texas A \& M University preprint CTP-TAMU-#1}

\nref\AN{R. Arnowitt and P. Nath, \PRL{69}{92}{725}; P. Nath and
R. Arnowitt, \PLB{287}{92}{89} and \PLB{289}{92}{368}.}
\nref\LNZ{\JL, \DVN, and A. Zichichi, \PLB{291}{92}{255}.}
\nref\LNP{\JL, \DVN, and H. Pois, \TAMU{61/92} and CERN-TH.6628/92.}
\nref\EKN{J. Ellis, S. Kelley and D. V.  Nanopoulos, \PLB{249}{90}{441}.}
\nref\EKNb{J. Ellis, S. Kelley and D. V.  Nanopoulos,
\PLB{260}{91}{131}; P. Langacker and M.-X. Luo, \PRD{44}{91}{817};
F. Anselmo, L. Cifarelli, A. Peterman, and A. Zichichi, Nuovo Cim. {\bf104A}
(1991) 1817.}
\nref\others{J. Ellis, S. Kelley, and \DVN, \NPB{373}{92}{55} and
\PLB{287}{92}{95}; R. Barbieri and L. Hall, \PRL{68}{92}{752}; F. Anselmo,
L. Cifarelli, A. Peterman, and A. Zichichi, Nuovo Cim. {\bf105A} (1992) 581;
J. Hisano, H. Murayama, and T. Yanagida, \PRL{69}{92}{1014}.}
\nref\WSY{S. Weinberg, \PRD{26}{82}{287}; N. Sakai and T. Yanagida,
\NPB{197}{82}{533}.}
\nref\ENR{J. Ellis, \DVN, and S. Rudaz, \NPB{202}{82}{43};
B. Campbell, J. Ellis, and \DVN, \PLB{141}{84}{229}.}
\nref\EMN{K. Enqvist, A. Masiero, and \DVN, \PLB{156}{85}{209}.}
\nref\CAN{P. Nath, A. Chamseddine, and R. Arnowitt, \PRD{32}{85}{2348};
P. Nath and R. Arnowitt, \PRD{38}{88}{1479}.}
\nref\ACPZ{F. Anselmo, L. Cifarelli, A. Peterman, and A. Zichichi, Nuovo Cim.
{\bf105A} (1992) 1201, and references therein.}
\nref\ACZ{F. Anselmo, L. Cifarelli, and A. Zichichi, CERN-PPE/92-145 and
\hfuzz=30pt CERN/LAA/MSL/92-011 (July 1992) (to appear in Nuovo Cimento).}
\nref\ANnc{P. Nath and R. Arnowitt, \MODA{2}{87}{331}.}
\nref\BT{H. Baer and X. Tata, Florida preprint FSU-HEP-920907.}
\nref\LNW{\JL, \DVN, and X. Wang, in preparation.}
\nref\aspects{For a detailed description of this general procedure see \eg,
S. Kelley, \JL, \DVN, H. Pois, and K. Yuan, \TAMU{16/92} and CERN-TH.6498/92.}
\nref\Bethke{S. Bethke, Talk given at the XXVI International Conference on
High Energy Physics, Dallas, August 1992.}
\nref\PDG{Particle Data Group, \PRD{45}{92}{S1}.}
\nref\Langacker{P. Langacker and N. Polonsky, Univ. of Pennsylvania preprint
UPR-0513T (October 1992).}
\nref\HMY{J. Hisano, H. Murayama, and T. Yanagida, \PRL{69}{92}{1014} and
Tohoku University preprint TU-400 (1992).}
\nref\Barbieri{R. Barbieri, F. Caravaglios, M. Frigeni, and M. Mangano,
\NPB{367}{91}{28}.}

\nfig\I{The calculated values of the proton lifetime into $p\to\bar\nu K^+$
versus the lightest chargino (or second-to-lightest neutralino) mass for both
signs of $\mu$. Note that we have taken $\alpha_3+1\sigma$ in order to
maximize $\tau_p$. Note also that future proton decay experiments should be
sensitive up to $\tau_p\approx20\times10^{32}\y$.}
\nfig\II{The correlation between the lightest chargino (or second-to-lightest
neutralino) and the lightest Higgs
boson masses for both signs of $\mu$. The bands for low values of $m_h$
correspond to $\tan\beta=1.5,1.75$. The plus signs indicate points where
the branching ratio into three charged leptons for neutralino-chargino
hadro-production
becomes negligible due to the opening of the channel $\chi^0_2\to \chi^0_1h$.}

\Titlehh{\vbox{\baselineskip12pt
\hbox{CERN-TH.6716/92}
\hbox{CERN/LAA/92-021}
\hbox{CTP--TAMU--72/92}
\hbox{ACT--21/92}}}
{\vbox{\centerline{Proposed Tests for Minimal SU(5)}
\centerline{Supergravity at Fermilab, Gran Sasso,}
\centerline{SuperKamiokande, and LEP}}}
\centerline{JORGE~L.~LOPEZ$^{(a)(b)}$, D.~V.~NANOPOULOS$^{(a)(b)(c)}$,
H. POIS$^{(a)(b)}$, and A. ZICHICHI$^{(d)}$}
\smallskip
\centerline{$^{(a)}$\CTPa}
\centerline{\CTPb}
\centerline{$^{(b)}$\HARCa}
\centerline{\HARCb}
\centerline{$^{(c)}$\CERN}
\centerline{$^{(d)}${\it CERN, Geneva, Switzerland}}
\centerline{ABSTRACT}
A series of predictions are worked out in order to put the minimal $SU(5)$
supergravity model under experimental test. Using the two-loop gauge coupling
renormalization group equations, with the inclusion of supersymmetric threshold
corrections, we calculate a new value for the proton
decay rate in this model and find that SuperKamiokande and Gran Sasso should
see the proton decay mode $p\to\bar\nu K^+$ for most of the allowed parameter
space. A set of physically sensible
assumptions and the cosmological requirement of a not too young Universe give
us a rather restrictive set of allowed points in the parameter space, which
characterizes this model. This set implies the existence of interesting
correlations among various masses: either the lightest chargino and the
next-to-lightest neutralino are below $\approx100\GeV$ (and therefore
observable at the Tevatron) or the lightest Higgs boson is below
$\approx50\GeV$ (and therefore observable at LEP I-II). These tests
are crucial steps towards selecting the correct low-energy effective
supergravity model. We also comment on the compatibility of the model with
$\sin^2\theta_w(M_Z)$ measurements as a function of $\alpha_3(M_Z)$.
\bigskip
{\vbox{\baselineskip12pt
\hbox{CERN-TH.6716/92}
\hbox{CERN/LAA/92-021}
\hbox{CTP--TAMU--72/92}
\hbox{ACT--21/92}}}
\Date{October, 1992}

One of the more interesting problems in high-energy physics is to disentangle
the right model for the description of all particles and all interactions.
Recently several rather restrictive constraints on the minimal $SU(5)$
supergravity model have been pointed out \refs{\AN,\LNZ,\LNP}. Here we
continue the study of this model and determine a set of predictions which could
be experimentally verified  with existing colliders and detectors.
These predictions mainly concern the lightest chargino and the next-to-lightest
neutralino, the lightest Higgs boson, and the proton lifetime.

Unification of the Standard Model particle interactions at very high energies
into larger models requires the presence of low-energy supersymmetry to
avoid the notorious gauge hierarchy problem. Moreover, supergravity models
allow one to explicitly calculate the phenomenologically necessary soft
supersymmetry breaking terms which split the ordinary particles from their
supersymmetric partners. The recent LEP measurements of the low-energy
gauge couplings and their use to study gauge coupling unification \refs{\EKN,
\EKNb,\others} constitutes a nice example of the validity of this scenario. In
this paper we restrict ourselves to the minimal $SU(5)$ supergravity model.
Here the introduction of the new light supersymmetric degrees of freedom raises
the unification scale $M_U$ and makes the dimension-six-operator mediated
proton ``partial" lifetime ($p\to e^+\pi^0$) much longer than experimentally
required. However, dimension-five proton decay operators \WSY\ arise due to the
exchange of a heavy colored Higgs triplet supermultiplet $H_3$ and can easily
give unacceptable proton lifetimes \ENR, unless $M_{H_3}\gsim M_U$ and the
supersymmetric spectrum is not too light \refs{\EMN,\CAN}. In fact, demanding
$M_{H_3}<3M_U$ so that the Yukawa coupling generating the $H_3$ mass remains
perturbative \refs{\EMN,\AN} and the naturalness criterion $m_{\tilde q,\tilde
g}<1\TeV$, it has been shown \AN\
that the proton decay mode $p\to \bar\nu K^+$ is sufficiently suppressed only
if the squarks and sleptons are heavy, and the two lightest neutralinos and
the lightest chargino are much lighter.

It has also been shown \LNZ\ that for representative points in the proton-decay
allowed five-dimensional parameter space of the model, the relic abundance of
the lightest neutralino -- a stable particle in the minimal model with
$R$-parity conservation -- is in gross conflict with current cosmological
observations mainly due to a lack of efficient pair-annihilation channels. The
cosmologically allowed set of points  was subsequently determined in an
extensive search of the parameter space \LNP. This set of points
shows an experimentally interesting correlation: if the lightest Higgs boson
mass is above $\approx80\GeV$, then the lightest chargino and the
second-to-lightest neutralino masses are  below $\approx100\GeV$, thus making
the observability of at least one of these particles quite likely at LEPII.

In this Letter we present a refinement of one of the most important elements in
the proton lifetime calculation, by determining the unification mass $M_U$
using \ACPZ\ the two-loop renormalization group equations (RGEs) for the gauge
couplings including the threshold effects of the supersymmetric particles.
(In Ref. \LNP, $M_U$ was calculated using the one-loop gauge coupling RGEs
with a common supersymmetric threshold at $M_Z$.) The main effect of these
corrections is a well known systematic reduction \refs{\EKN,\ACZ} of the value
of $M_U$ which in turn reduces the upper bound on the proton
lifetime (since $\tau_p\propto M^2_{H_3}$ and we take $M_{H_3}<3M_U$) rendering
most of the originally allowed points unacceptable. We also explore some
previously neglected regions of parameter space (where $\tan\beta<2$) which
maximize the proton lifetime and increase the number of allowed points. The
final allowed set of points entails a limit on the lightest chargino and
second-to-lightest neutralino masses of $m_{\chi^\pm_1,\chi^0_2}\lsim150\GeV$
and on the lightest Higgs bosons mass of $m_h\lsim100\GeV$.
Since neutralino-chargino hadro-production at Fermilab should be able to probe
$m_{\chi^\pm_1,\chi^0_2}\lsim100\GeV$ through the tri-lepton signal
\refs{\ANnc,\BT,\LNW}, most of the parameter space of this model can be
probed in the near future. Furthermore, if
$m_{\chi^\pm_1,\chi^0_2}\gsim100\GeV$, then
$m_h\lsim50\GeV$, making observability of the latter at LEP a sure bet.

Let us first sketch the calculational procedure. The low-energy
sector of the minimal $SU(5)$ supergravity model can be described in terms of
five parameters: the universal soft-supersymmetry breaking scalar ($m_0$) and
gaugino ($m_{1/2}$) masses and the trilinear coupling ($A$), the ratio of
vacuum expectation values ($\tan\beta$), and the top-quark mass ($m_t$).
This reduced set of parameters results from the use of the RGEs to relate
high- and low-energy parameters and the requirement of radiative breaking
of the electroweak symmetry (enforced by using the one-loop effective
potential) \aspects. For quick reference, $m_{1/2}$ determines the gluino mass
$m_{\tilde g}=(\alpha_3/\alpha_U) m_{1/2}$, while $m_0$ and $m_{1/2}$ determine
the squark and slepton masses $m^2_{\tilde q,\tilde l}\approx
m^2_0+c_im^2_{1/2}$, with $c_i\sim6\,(0.5)$ for squarks (sleptons). The Higgs
mixing parameter $\mu$ is calculable, it roughly scales with $m_0$ or $m_{1/2}$
(whichever is dominant) and grows with increasing $m_t$; its sign remains
undetermined.

In Ref. \LNP\ three of us conducted a discrete scan of this five-dimensional
parameter space and determined those points which satisfied all experimental
bounds on the sparticle and one-loop corrected Higgs boson masses. We used
one-loop gauge coupling RGEs and a common supersymmetric threshold at $M_Z$, to
determine $M_U,\alpha_U$, and $\sin^2\theta_w$, once
$\alpha_3(M_Z)=0.118\pm0.008$ \Bethke\ and $\alpha^{-1}_e(M_Z)=127.9$ were
given. We then computed the proton lifetime into the
dominant decay mode $p\to\bar\nu_{e,\mu,\tau}K^+$, assuming $M_{H_3}<3M_U$,
and obtained an upper bound on $\tau_p$. About $10\%$ of the otherwise
allowed points satisfied $\tau_p>\tau^{exp}_p=1\times10^{32}\y$ \PDG. We also
computed the relic abundance of the lightest neutralino $\Omega_\chi h^2_0$,
where $\Omega_\chi=\rho_\chi/\rho_0$ is the present neutralino density relative
to the critical density and $0.5\le h_0\le1$ is the scaled Hubble parameter.
Only about $1/6$ of the proton-decay allowed points satisfied the
observationally required bound $\Omega_\chi h^2_0\le1$.

Our refined study here
includes several new and important features: (i) recalculation of $M_U$ using
two-loop gauge coupling RGEs including light supersymmetric thresholds, (ii)
exploration of values of $\alpha_3$ throughout its $\pm1\sigma$ allowed range,
and (iii) exploration of low values of $\tan\beta\,(<2)$ which maximize
$\tau_p$. We use the analytical approximations to the solution of the two-loop
gauge coupling RGEs in Ref. \ACPZ\ to obtain $M_U,\alpha_U,\sin^2\theta_w$. The
supersymmetric threshold is treated in great detail \ACPZ\ with all the
sparticle masses obtained from our procedure \LNP. Since, as discussed in the
previous paragraph, these vary throughout the allowed parameter space, we
obtain {\it ranges} for the calculated values. In Table I we show the one-loop
value for $M_U$ ($M_U^{(0)}$), the two-loop plus supersymmetric threshold
corrected unification mass range ($M_U^{(1)}$) [as expected \ACPZ\ $M_U$ is
reduced by both effects], the ratio of the two, and the calculated range of
$\sin^2\theta_w$. These values are obtained after all constraints have been
satisfied, the proton decay being the most important one. Note that for
$\alpha_3=0.118$ (and lower), $\sin^2\theta_w$ is outside the experimental
$\pm1\sigma$ range ($\sin^2\theta_w=0.2324\pm0.006$ \Langacker), whereas
$\alpha_3=0.126$ gives quite acceptable values.
\topinsert
\hrule\smallskip
\noindent{\bf Table I}: The value of the one-loop unification mass $M_U^{(0)}$,
the two-loop and supersymmetric threshold corrected unification mass range
$M_U^{(1)}$, the ratio of the two, and the range of the calculated
$\sin^2\theta_w$, for the indicated values of $\alpha_3$ (the superscript
$+\,(-)$ denotes $\mu>0\,(<0)$) and $\alpha^{-1}_e=
127.9$. The $\sin^2\theta_w$ values should be compared with the current
experimental $\pm1\sigma$ range $\sin^2\theta_w=0.2324\pm0.0006$ \Langacker.
Lower values of $\alpha_3$ drive $\sin^2\theta_w$ to values even higher than
for $\alpha_3=0.118$. All masses in units of $10^{16}\GeV$.
\smallskip
\input tables
\thicksize=1.0pt
\centerjust
\begintable
\|$\alpha_3=0.126^+$|$\alpha_3=0.126^-$\|$\alpha_3=0.118^+$|$\alpha_3=0.118^-$
\cr
$M^{(0)}_U$\|$3.33$|$3.33$\|$2.12$|$2.12$\nr
$M^{(1)}_U$\|$1.60-2.13$|$1.60-2.05$\|$1.02-1.35$|$1.02-1.30$\nr
$M_U^{(1)}/M_U^{(0)}$\|$0.48-0.64$|$0.48-0.61$\|$0.48-0.64$|$0.48-0.61$\nr
$\sin^2\theta_w$\|$0.2315-0.2332$|$0.2313-0.2326$\|$0.2335-0.2351$|
$0.2332-0.2345$\endtable
\smallskip
\hrule\medskip
\endinsert

We do not specify the details of the GUT thresholds and in practice take the
usual three GUT mass parameters ($M_V,M_\Sigma,M_{H_3}$) to be degenerate with
$M_U$. Since we then allow $M_{H_3}<3M_U$, Table I indicates that in our
calculations $M_{H_3}<6.4\times10^{16}\GeV$. In Ref. \HMY\ it is argued that a
more proper upper bound is $M_{H_3}<2M_V$, but $M_V$ cannot be calculated
directly, only $(M^2_V M_\Sigma)^{1/3}<3.3\times10^{16}\GeV$ is known from
low-energy data \HMY. If we take $M_\Sigma=M_V$, this would give
$M_{H_3}<2M_V<6.6\times10^{16}\GeV$, which agrees with our present requirement.
Below we comment on the case $M_\Sigma<M_V$.

With the new value of $M_U$ we  simply rescale our previously calculated
$\tau_p$ values which satisfied $\tau^{(0)}_p>\tau^{exp}_p$, and find that
$\tau^{(1)}_p=\tau^{(0)}_p[M_U^{(1)}/M_U^{(0)}]^2<\tau^{exp}_p$ for $\gsim75\%$
of them. The value of $\alpha_3$ has a significant influence on the results
since (see Table I) larger (smaller) values of $\alpha_3$ increase (decrease)
$M_U$, although the effect is more pronounced for low values of $\alpha_3$. To
quote the most conservative values of the observables, in what follows we take
$\alpha_3$ at its $+1\sigma$ value ($\alpha_3=0.126$). As discussed above, this
choice of $\alpha_3$ also gives
$\sin^2\theta_w$ values consistent with the $\pm1\sigma$ experimental range.
Finally, in our previous search \LNP\ of the parameter space we considered only
$\tan\beta=2,4,6,8,10$ and found that $\tan\beta\lsim6$ was required. Our
present analysis indicates that this upper bound is reduced down to
$\tan\beta\lsim3.5$. Here we consider also $\tan\beta=1.5,1.75$ since low
$\tan\beta$ maximizes $\tau_p\propto \sin^22\beta$. These add new allowed
points (\ie, $\tau^{(0)}_p>\tau^{exp}_p$) to our previous set, although most of
them ($\gsim75\%$) do not survive the stricter proton decay constraint
($\tau^{(1)}_p>\tau^{exp}_p$) imposed here.

In Fig. 1 we show the re-scaled values of $\tau_p$ versus the lightest
chargino mass $m_{\chi^\pm_1}$. All points satisfy $\xi_0\equiv
m_0/m_{1/2}\gsim6$ and $m_{\chi^\pm_1}\lsim150\GeV$, which are to be contrasted
with $\xi_0\gsim3$ and $m_{\chi^\pm_1}\lsim225\GeV$ derived in Ref. \LNP\ using
the weaker proton decay constraint. The upper bound
on $m_{\chi^\pm_1}$ derives from its near proportionality to $m_{\tilde g}$,
$m_{\chi^\pm_1}\approx0.3m_{\tilde g}$ \refs{\AN,\LNP}, and the result
$m_{\tilde g}\lsim500\GeV$. The latter follows from the proton decay constraint
$\xi_0\gsim6$ and the naturalness requirement $m_{\tilde q}\approx\sqrt{m^2_0+
6m^2_{1/2}}\approx{1\over3}m_{\tilde g}\sqrt{6+\xi^2_0}<1\TeV$. Within our
naturalness and $H_3$ mass assumptions, we then obtain \foot{Note that in
general, $\tau_p\propto M^2_{H_3}[m^2_{\tilde q}/
m_{\chi^\pm_1}]^2\propto M^2_{H_3}[m_{\tilde g}(6+\xi^2_0)]^2$ and thus
$\tau_p$ can be made as large as desired by increasing sufficiently either
the supersymmetric spectrum or $M_{H_3}$.}
\eqn\I{\tau_p<3.1\,(3.4)\times10^{32}\y\quad{\rm for}\quad \mu>0\,(\mu<0).}
The $p\to\bar\nu K^+$ mode should then be readily observable at SuperKamiokande
and Gran Sasso since these experiments should be sensitive up to
$\tau_p\approx2\times10^{33}\y$. Note that if $M_{H_3}$ is relaxed up to its
largest possible value consistent with low-energy physics,
$M_{H_3}=2.3\times10^{17}\GeV$ \HMY, then in Eq. \I\ $\tau_p\to
\tau_p<4.0\,(4.8)\times10^{33}\y$, and only part of the parameter space of
the model would be experimentally accessible. However, to make this choice
of $M_{H_3}$ consistent with high-energy physics (\ie, $M_{H_3}<2M_V$) one must
have $M_V/M_\Sigma>42$.

In this model the only light particles are the lightest Higgs boson
($m_h\lsim100\GeV$), the two lightest neutralinos ($m_{\chi^0_1}\approx
{1\over2}m_{\chi^0_2}\lsim75\GeV$), and the lightest chargino ($m_{\chi^\pm_1}
\approx m_{\chi^0_2}\lsim150\GeV$). The gluino and the lightest stop can be
light ($m_{\tilde g}\approx 160-460\GeV$, $m_{\tilde t_1}\approx170-825\GeV$),
but for most of the parameter space are not within the reach of Fermilab. In
Fig. 2 we present the results for the experimentally interesting
correlation between $m_h$ and $m_{\chi^\pm_1}\approx m_{\chi^0_2}$, which
shows\foot{In Fig. 2 only $\tan\beta=1.5,1.75,2$ are shown. For the maximum
allowed $\tan\beta$ value ($\approx3.5$), $m_h\lsim100\GeV$.} that
$m_h\lsim80\,(95)\GeV$ for $\mu>0\,(\mu<0)$. The bands of points towards low
values of $m_h$ represent the discrete choices of $\tan\beta=1.5,1.75$. The
voids between these bands are to be understood as filled by points with
$1.5\lsim\tan\beta\lsim1.75$. For
arbitrary values of the chargino or next-to-lightest neutralino masses, almost
all of the Higgs mass range should be explorable at LEPII. Moreover, for
$m_{\chi^\pm_1}>106\,(92)\GeV$ (for $\mu>0\,(\mu<0)$), we obtain
$m_h\lsim50\GeV$ and Higgs detection at LEP should be immediate.\foot{In Fig. 2
for $\mu>0$, $m_h\approx50\GeV$, and $m_{\chi^\pm_1}\gsim100\GeV$, there is
a sparsely populated area with highly fine-tuned points in parameter space
($m_t\approx100\GeV$, $\tan\beta\approx1.5$, $\xi_A\approx\xi_0\approx6$).}
This updated prediction ($m_h\gsim50\GeV\Rightarrow
m_{\chi^\pm_1,\chi^0_2}\lsim100\GeV$) is much sharper that the previous one
($m_h\gsim80\GeV\Rightarrow m_{\chi^\pm_1,\chi^0_2}\lsim100\GeV$) in Ref. \LNP.

Interestingly enough, it has been recently pointed out \BT\ that Fermilab
has the potential of exploring most of the LEPII parameter space, before
LEPII turns on. This would occur through the process $p\bar p\to
\chi^0_2\chi^\pm_1$ which has a cross section $\gsim1\pb$ for
$m_{\chi^\pm_1,\chi^0_2}
\lsim100\GeV$. The further decay into three charged leptons has very little
background \refs{\ANnc,\Barbieri,\BT} and possibly sizeable branching ratios
which, with an integrated luminosity of $\approx100\ipb$, should yield a
significant number of candidate events. A detailed calculation of this process
in this model is in progress \LNW. One concern which is usually brushed aside
is whether the decay channel $\chi^0_2\to \chi^0_1h$ is open, since in this
case the branching ratio into three charged leptons is expected to be
negligible. We have checked that in this model this channel
is indeed open, although mostly for $m_{\chi^\pm_1,\chi^0_2}\gsim100\GeV$, and
thus
mostly inconsequential for Fermilab searches. These points are represented
by pluses in Fig. 2 and, given the approximate mass relations in this model,
roughly speaking correspond to $m_h\lsim{1\over2}m_{\chi^\pm_1,\chi^0_2}$.

In sum, within our physically reasonable assumptions, the minimal $SU(5)$
supergravity model should be fully testable at a combination of present and
near future experimental facilities. At Fermilab, chargino-neutralino
hadro-production and decay into three charged leptons should probe
$m_{\chi^\pm_1,\chi^0_2}
\lsim100\GeV$. If this process is not observed at Fermilab, then LEP should
see the lightest Higgs below $\approx50\GeV$. If Fermilab does see the
chargino-neutralino, then LEPII should confirm the model by observing the Higgs
at the
appropriate mass. Independently, SuperKamiokande and Gran Sasso should
see the $p\to\bar\nu K^+$ decay mode for most of the allowed parameter space.
Of course, if any of the above predictions fails to be confirmed, then under
our physically reasonable assumptions the minimal $SU(5)$ supergravity model
will be excluded.
\bigskip
\bigskip
\noindent{\it Acknowledgments}: This work has been supported in part by DOE
grant DE-FG05-91-ER-40633. The work of J.L. has been supported by an SSC
Fellowship. The work of D.V.N. has been supported in part by a grant from
Conoco Inc. We would like to thank the HARC Supercomputer Center for the use
of their NEC SX-3 supercomputer.
\listrefs
\listfigs
\bye